\documentclass[11pt]{iopart}
\usepackage{color}
\usepackage{cite}
\usepackage[colorlinks=true,linkcolor=blue,urlcolor=blue,citecolor=blue]{hyperref}
\usepackage{upgreek}
\usepackage{footmisc}
\usepackage[colorinlistoftodos]{todonotes}

\begin{document}

\title[]{Precision Control of Resistive Power in Kibble Balance Coils: An Advanced Method for Minimizing Temperature-Related Magnetic Errors}

\author{Weibo Liu$^1$, Stephan Schlamminger$^2$, Shisong Li$^{1}$}

\address{1. Department of Electrical Engineering, Tsinghua University, Beijing 100084, China\\
2. National Institute of Standards and Technology (NIST), Gaithersburg, 20899 MD, USA}
\ead{shisongli@tsinghua.edu.cn}
\vspace{10pt}

\begin{abstract}
Temperature changes affect the coercivity of permanent magnets, thereby impacting the $Bl$ factor and potentially introducing systematic errors in Kibble balance measurements. While the thermal-magnetic effect is negligible in large magnet systems, it increases substantially as the magnet size decreases, posing an engineering difficulty for tabletop Kibble balance systems. We discuss the mechanism of thermal-magnetic effects through finite element analysis, which has not been sufficiently emphasized in previous studies. A bifilar-coil power regulator is proposed to eliminate thermal-magnetic errors in Kibble balances. The approach aims to keep the power of the internal heating source—coil ohmic power—constant over time, allowing the $Bl$ drift to be mitigated through ABA or ABBA measurement sequences. Experimental results validate the proposal, demonstrating that the thermal effect can be reduced by more than two orders of magnitude compared to the conventional two-mode, two-phase measurement scheme, and by about one order of magnitude compared to the one-mode, two-phase scheme. The proposed approach can eliminate the influence of thermal-magnetic effects on the measurement results, thus further breaking down the limitations on the minimum size of tabletop Kibble balances.
\end{abstract}




\clearpage
\begin{flushleft}

\section{Introduction}

The Kibble balance~\cite{Kibble1976} is one of the two primary methods for mass realization in the new International System of Units (SI) \cite{RevisedSI2019}. Nowadays, many national metrology institutes (NMIs) \cite{Fang_2020BIPM,NIST2017,NRC2017,NIM2023,METAS2022,Thomas_2017LNE,KRISS2020,MSL2020,UME2023} have carried out studies on Kibble balances and achieved impressive progress. The fundamental principles and experimental details of a typical Kibble balance can be found in related review articles, such as \cite{Robinson_2016watt}.

A typical Kibble balance operates through two measurement phases: the weighing phase and the velocity phase. In the weighing phase, the Lorentz force generated by the coil in the magnetic field balances the gravitational force acting on the test mass, expressed as $mg = (Bl)_{\rm{w}}I$, { where $m$ is the mass of
the test artefact}, and $g$ is the local gravitational acceleration and $(Bl)_{\rm{w}}$ is the magnetic geometrical factor, $B$ is the magnetic flux density at the coil position, $l$ is the coil wire length, and the subscript $_{\rm{w}}$ indicates the $B$ and $l$ product is evaluated in the weighing phase. In the velocity phase, the coil is disconnected from the circuit and moves vertically at a constant speed $v$ within the same magnetic field, inducing a voltage $U = (Bl)_{\rm{v}}v$. When the equivalence $(Bl)_{\rm{w}} = (Bl)_{\rm{v}}$ is ensured, the mass can be realized as $m = {UI}/{(gv)}$. By utilizing quantum electrical standards, such as the Josephson voltage standard and the quantum Hall resistance standard~\cite{JOSEPHSONeffect,Halleffect}, as well as time and length standards, the precise link between electrical power and mechanical power can be established \cite{Haddad_Bridging2016} and the mass can be determined with a relative measurement uncertainty of a few parts in $10^8$ at the kilogram level.

High-precision Kibble balances, such as~\cite{NISTInvitedArticle2016,NRC2017}, are typically large in size. In recent years, however, many research groups and industrial stakeholders have shown significant interest in developing smaller, more cost-effective, and easier-to-operate Kibble balance systems, particularly popular are tabletop systems ~\cite{NISTtabletop2020,NIST_QEMMS,PTB_PB2_2021,NPLtabletop2024}. One challenge with miniaturizing the Kibble balance is that some systematic effects are related to the size of the magnet and they tend to get worse as the magnet is made smaller. Preliminary studies~\cite{Li_MagneticUncertainties} have already demonstrated that certain magnetic errors become more pronounced when using smaller magnets, including the coil current effect~\cite{CoilCurrent_Li2018}, magnetic nonlinearity~\cite{hysteresis2020,Nonlinear_2014,Nonlinear_2013}, and the thermal-magnetic effect.
Among these, the thermal-magnetic effect is the primary limitation impeding the miniaturization of Kibble balance magnets. According to \cite{Li_MagneticUncertainties}, if a linear scale factor $p$ (typically $0<p<1$) is used to represent the ratio of the one-dimensional size of a Kibble balance magnet system compared to a reference magnet, e.g.~\cite{NISTshimming}, the relative measurement uncertainty introduced by the thermal-magnetic effect scales approximately as $p^{-8}$. Therefore, suppressing the thermal-magnetic effect is essential for achieving high accuracy in a tabletop Kibble balance.

To eliminate the thermal-magnetic errors in tabletop Kibble balances, we analyze the measurement sequences of several typical Kibble balance schemes and investigate the underlying causes of these errors.
We propose a novel approach to regulate the coil's resistive power loss. This effectively reduces temperature variations during the measurement sequence, {as well as the current polarity change during the weighing measurement}, and thus further breaks down the limitations imposed by thermal-magnetic effects on the minimum size of tabletop instruments. 
Note that this work is an addition to the recent updates of the Tsinghua tabletop Kibble balance project~\cite{THUdesign2022,THUKB2023, li2024magnet, THU2024currentsource, THU2024gravity}, aiming to construct a compact, highly accurate, user-friendly, open-hardware, and cost-effective Kibble balance instrument. 

The article is organized as follows: Section \ref{section2} reviews thermal-magnetic errors, explores their causes across different Kibble balance measurement schemes, and proposes a new thermal control scheme by regulation of the coil ohmic power. Its effectiveness is verified through numerical analysis in section \ref{section3}, and the thermal-magnetic error with different measurement schemes is evaluated. An experimental test to verify the proposal is presented in section \ref{section4}. Section \ref{section5} discusses the practical application of the proposed approach, including the influence of the measurement period on thermal effects and the zero-thermal-effect point for the conventional two-mode two-phase
measurement scheme, the overshoot in the Lorentz force when conducting the new measurement scheme, and the determination of tabletop Kibble balances' size boundaries considering thermal-magnetic errors. The article ends with a conclusion  in section \ref{section6}.

\section{Principle of the proposed coil ohmic power regulator}
\label{section2}

The thermal-magnetic error in Kibble balance systems arises from the temperature-dependent coercivity of the permanent magnets used to generate $B$. The temperature coefficient $\alpha_{\rm{m}}$ is commonly employed to quantify the temperature dependence of a permanent magnet \cite{Li_2022_irony}. For example, Samarium-Cobalt (SmCo) magnets have a temperature coefficient of approximately $-3\times 10^{-4}$/K, indicating that a 1\,K increase in temperature results in a relative decrease of $3 \times 10^{-4}$ in the air-gap flux density $B$. Some research groups use Neodymium-Iron-Boron (NdFeB) as a magnet material, such as in \cite{li2024magnet}, which can generate a stronger magnetic field but have a larger temperature coefficient of about $-1 \times 10^{-3}$/K. Thermal-magnetic errors occur when there is a temperature difference between the weighing and velocity phases, which disrupts the equivalence condition $(Bl)_{\rm{w}} = (Bl)_{\rm{v}}$ and introduces a measurement bias in the final mass determination. The true mass result $m'$ in contrast to the measured result would be
\begin{equation}
	m'=\frac{UI}{gv}\frac{(Bl)_\mathrm{w}}{(Bl)_\mathrm{v}}.
\end{equation}

The measurement result, $m=UI/(gv)$, compared to the true value $m'$ changes relatively by
\begin{equation}
	\epsilon=\frac{\Delta m}{m'}\approx\frac{m-m'}{m}=1-\frac{(Bl)_\mathrm{w}}{(Bl)_\mathrm{v}}=\alpha_\mathrm{m}(T_\mathrm{v}-T_\mathrm{w}),
\end{equation}
where $T_\mathrm{w}$ and $T_\mathrm{v}$ denote the average temperature of the permanent magnet during the weighing and velocity measurements, respectively. We call $\epsilon$ the thermal-magnetic error.

To minimize thermal-magnetic errors, experimenters can either reduce the temperature coefficient of the permanent magnet material $\alpha_\mathrm{m}$ or stabilize the temperature during measurements $T_\mathrm{v}-T_\mathrm{w}$. One approach, proposed by researchers at METAS (Federal Institute of Metrology, Switzerland) and used by the Kibble balance groups at NIST (National Institute of Standards and Technology, USA), involves using SmCoGd (Samarium-Cobalt-Gadolinium) magnets as the permanent magnet material \cite{METAS_Baumann_2013, NIST_QEMMS}. The temperature coefficient of SmCoGd is approximately $-1 \times 10^{-5}$/K \cite{METAS7390181}, which is significantly lower than that of SmCo and NdFeB magnets. However, this reduction in temperature sensitivity comes at the cost of a 20\,\% to 30\,\% decrease in coercivity, and, hence, reduced $Bl$ for a given geometry, compared to SmCo magnets \cite{Robinson_2016watt, Li_2022_irony}.

An alternative method for reducing thermal sensitivity is to optimize the magnetic circuit through a technique known as magnetic-shunt compensation, as implemented in the METAS Mark II system \cite{METAS_Baumann_2013, METAS7390181}. This method involves using a temperature-sensitive Fe–Ni alloy as a magnetic shunt. Due to its strong temperature dependence and low Curie temperature, the Fe–Ni alloy compensates for temperature-induced changes in the main flux path, thereby reducing the overall temperature dependence of the magnetic flux density. By combining SmCoGd magnets with a carefully tuned Fe–Ni shunt, METAS has achieved a temperature coefficient as low as $10^{-6}$/K \cite{METAS_Baumann_2013, METAS7390181}. 
Unfortunately, effective compensation is difficult to achieve for magnets with higher temperature coefficients, such as NdFeB. Additionally, thermal gradients between the magnet and the shunt can lead to spatially varying compensation, making the method less practical in systems with significant thermal variations.

Another method to mitigate thermal-magnetic effects is to minimize temperature variations in the magnet system. One straightforward solution is to implement an environmental temperature control system. For instance, the NPL/NRC Kibble balance employs an indoor air-conditioning system with temperature feedback for precise environmental control \cite{Robinson_2012NPL}. This setup has demonstrated a temperature drift of the magnet as low as $-400\,\upmu\rm{K}$/day. While temperature-controlled ovens are highly effective at eliminating external environmental fluctuation, their performance can degrade when internal heat sources, such as resistive heating in the coil, are present. The placement of the temperature sensor is also of importance. While it is possible to control a single point, gradients can still arise in the system. Servo controlling a single point temperature of the magnet does by no means correspond to controlling the $Bl$.
\begin{figure}
	\centering
	\includegraphics[width=0.8\columnwidth]{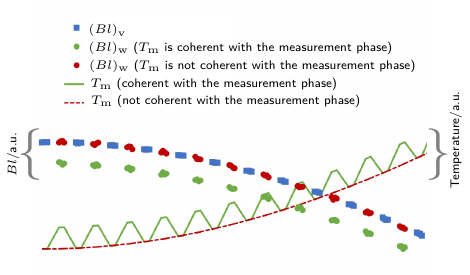}
	\caption{A schematic of the thermal magnetic error for a typical Kibble balance. 
		The horizontal axis denotes the measurement time, and $T_\mathrm{m}$ represents the temperature of the permanent magnet. The green solid curve shows a temperature variation that is coherent with the phase of the Kibble experiment, i.e., higher temperature during the weighing phase, which in turn leads to a smaller $Bl$ for the weighing measurement due to the thermal-magnetic effect. The red dashed curve shows a drift of the temperature that is independent of the Kibble balance phase. In both cases, the $Bl$ change related to the slow drift part of $T_\mathrm{m}$ can be extracted by a polynomial fit, detailed in section \ref{section 3.1}.}
	\label{fig:01}
\end{figure}

As illustrated in Figure \ref{fig:01}, a typical Kibble balance alternates between weighing and velocity measurements. This measurement sequence allows the application of the ABA or ABBA scheme \cite{Swanson_2010ABAB} or polynomial fitting to effectively mitigate systematic effects arising from environmental temperature drift. Consequently, any gradual thermal changes experienced during both the weighing and velocity phases will not produce systematic errors. While the environmental temperature control discussed in \cite{Robinson_2012NPL} may offer some advantages, it is important to note that environmental temperature drifts and other slow temperature variations unrelated to the measurement sequence do not impact the final results of the Kibble balance.

Although the ABA and ABBA schemes can address slow temperature changes, they can not suppress effects due to temperature variations that are coherent with the measurement process itself, particularly the effects caused by coil ohmic power. In fact, the temperature fluctuations associated with the measurement sequence, present the most significant and complex challenges to accurate measurement results \cite{Robinson_2016watt,Li_MagneticUncertainties}. Using the measurement process of a typical Kibble balance NIST-4 as an example, the alternating velocity and weighing measurements aim to mitigate the effects of slow temperature drift by the ABA scheme. However, since the coil current is switched off during the velocity phase and then turned on for the weighing phase, the periodic alteration of the coil's heating state causes temperature inconsistencies in the magnets across both phases, as depicted in Figure \ref{fig:01}. This results in a systematic thermal-magnetic error that can lead to a systematic bias in the measurement result~\cite{Li_MagneticUncertainties}.

We examine existing typical Kibble balance measurement schemes to effectively analyze the thermal-magnetic effects coherent with the measurement sequence caused by coil ohmic power. This analysis will help identify optimization methods that can mitigate these thermal-magnetic effects.

Currently, two measurement schemes are commonly employed in typical Kibble balances: the two-mode, two-phase (TMTP) measurement scheme, and the one-mode, two-phase (OMTP) measurement scheme \cite{Fang_2020BIPM,THUdesign2022}. The traditional TMTP scheme is used by the vast majority of Kibble balances nowadays, typically the NIST-4 \cite{NIST2017} and the NPL/NRC Kibble balance \cite{NRC2017}, and OMTP scheme is mainly adopted by the BIPM Kibble Balance \cite{Fang_2020BIPM}. As defined in \cite{Robinson_2016watt}, here "phase" refers to the temporal relationship between the velocity and weighing measurements. In both of the two-phase schemes (TMTP and OMTP), these processes occur in separate time intervals. "Mode" denotes the number of times the coil current state is switched during a complete measurement cycle. In the TMTP scheme, which utilizes a single coil, the current is switched on during the weighing phase and off during the velocity phase, thus involving two modes. Conversely, in the OMTP scheme, which employs a bifilar coil (consisting of coil \#1 and coil \#2), the current remains on during both the weighing and velocity phases, representing a single mode.
\begin{figure}[tp!]
	\centering
	\includegraphics[width=0.8\columnwidth]{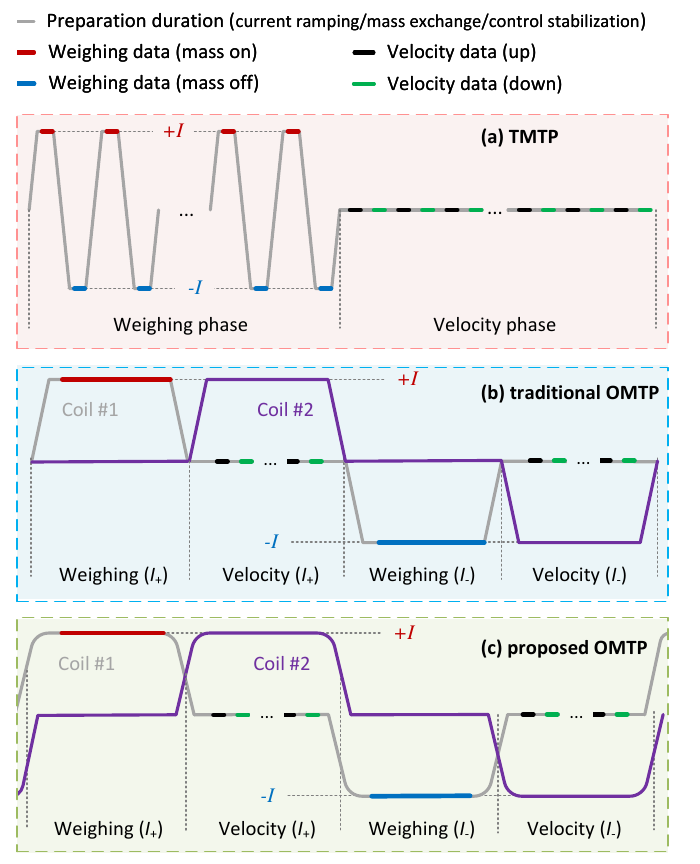}
	\caption{Full measurement period for three current ramping strategies. (a), (b) and (c) present the TMTP measurement scheme, the traditional OMTP scheme, and the proposed scheme,  respectively. The horizontal axis spans one measurement cycle in time, and the vertical axis shows the normalized coil current.}
	\label{fig:02}
\end{figure}

Figure \ref{fig:02} (a) and (b) show the coil currents in typical TMTP and OMTP measurement sequences. For both TMTP and OMTP measurement schemes, thermal-magnetic effects arise due to the ohmic power dissipation in the coil. However, the underlying mechanisms causing these thermal-magnetic errors differ between the two schemes. In the TMTP scheme, the thermal-magnetic effect is induced by the periodic switching of the coil current between the weighing and velocity phases, which results in a temperature difference $\Delta T_\mathrm{m}=T_{\rm{v}}-T_{\rm{w}}$ and subsequently a corresponding difference in the magnetic geometrical factor, $(BL)_{\rm{v}}-(BL)_{\rm{w}}$, between two phases. 

In contrast, the traditional OMTP scheme maintains a constant coil current throughout both the weighing and velocity phases, thereby reducing the magnitude of thermal-magnetic errors compared to the TMTP scheme. However, the conventional OMTP scheme includes a setup time before each weighing or velocity measurement. This time is used for current ramping, mass exchange, and control stabilization. In most cases, a linear ramp for the current is used and the coils are switched when the current in both coils is zero. During the ramp, the ohmic power of the double coils is lower than during the actual measurement, leading to temperature fluctuations associated with the measurement sequence, which in turn introduces system errors.

We propose a new measurement scheme based on the OMTP approach, as illustrated in Figure \ref{fig:02} (c). The key idea is to introduce a second current source that tracks the current and power in the measurement coil (in this case, coil \#1 in Figure \ref{fig:02}) and feeds current back into the other coil (coil \#2) to maintain a constant total ohmic power in the bifilar coil system. Let the current through coil \#1 and coil \#2 be $i_1(t)$, $i_2(t)$, and the resistance be $R_1$, $R_2$, then the total power of the bifilar coils $P_\mathrm{c}(t)$ is written as
\begin{equation}
	P_\mathrm{c}(t)=i_1^2(t) R_1+i_2^2(t) R_2.
	\label{Pc_equation1}
\end{equation}

To keep a constant total ohmic power $P_\mathrm{c}$, the coil \#2 current is set according to the current of coil \#1, i.e.
\begin{equation}
	i_2(t)=\sqrt{\frac{P_\mathrm{c}-i_1^2(t)R_1}{R_2}}.
	\label{Pc_equation2}
\end{equation}

By implementing this active power control, $P_\mathrm{c}(t)$ remains constant throughout the entire experiment. As mentioned above, the magnetic field variations caused by gradual temperature changes can be effectively addressed using ABA or ABBA measurement sequences, or by applying polynomial fitting. So by controlling the dissipative power, no systematic temperature differences are observed between different measurement phases, thus eliminating all thermal-magnetic errors.

It is important to note that the proposed ohmic power compensation should only be applied during the current ramps, and the compensation current must be turned off during the measurement period. This is because current fluctuations in the second coil could introduce additional measurement errors, either by directly affecting the force measurement or inducing voltage due to current variations \cite{CoilCurrent_Li2018}. A complete turn off can be easily implemented with a switch box. With this approach, there is no interference on the primary measurement on coil \#1 during measurement, and the only change is that coil \#2 now serves as an auxiliary heating coil to regulate total ohmic power during current ramping. The only disadvantage is added complexity as the power compensation requires a second current source. However, since the power control does not require high precision, the additional cost associated with the second current source would be minimal.

In the following, we introduce a simple design on the ramping of current of $i_1$ and $i_2$, which will be used in the Tsinghua tabletop Kibble balance. When the current in the main coil ramps starting from zero with the target current $I_1$, $i_1$ and $i_2$ are set as 
\begin{equation}
	i_1(t)=I_1\sin\left(\frac{\pi}{2\tau_{\rm{r}}}(t-t_0)\right) \;\;\mbox{and}
	\label{i1_v1_eq}
\end{equation}
\begin{equation}
	i_2(t)=I_2\cos\left(\frac{\pi}{2\tau_{\rm{r}}}(t-t_0)\right),
	\label{i2_v1_eq}
\end{equation}
where $\tau_{\rm{r}}$ is the time designed for $i_1$ ramping from 0 to $I_1$, $t_0$ is the ramping start time, $I_2$ is the current $i_2$ at $t_0$. 
For turning the currents in coil \#1 off the sine and cosine are interchanged, $i_1$ changes like a cosine from $I_1$ to 0 and $I_2$ like a sine from 0 to $I_2$.

In the Tsinghua tabletop Kibble balance, a bifilar coil containing two identical independent coils is used. In this case, the coil resistance $R_1\approx R_2=R$. Using the above current setup and a symmetrical current during mass-on and mass-off, $|I_1|=|I_2|=I$, the total power is kept constant as that during the whole experimental process, i.e.
\begin{equation}
	P_\mathrm{c}=I^2R.
\end{equation}

\section{Analysis of the thermal-magnetic error with finite element analysis}
\label{section3}

The thermal-magnetic error is evaluated using finite element analysis (FEA). The objective is to analyze temperature variations during the measurement sequence in TMTP, OMTP, and the proposed method. Particularly, we are interested in how effective the proposed thermal control is in reducing measurement bias due to the thermal-magnetic effects. For the following calculations, we use the NIST-4 magnet system \cite{NISTshimming} as a reference, whose outer diameter of the magnet system is 600\,mm, i.e., the scale factor $p$ introduced in~\cite{Li_MagneticUncertainties} is set to one. {In order to simplify the FEA modeling, the following assumptions are made:.

\begin{itemize}
\item Only the permanent magnet (SmCo or NdFeB), yoke (iron), and coil (copper) are included in the model. Other components, such as the coil suspension, are not modeled.
\item The FEA model is scaled uniformly as the scale factor $p$ changes.
\item The position of the coil is fixed, disregarding the motion of the coil during the velocity phase.
\item The FEA simulation accounts for thermal conduction and convection in air, but radiation is not modeled.
\item Thermal expansion of solid materials is neglected in the analysis.
\end{itemize}

}

To allow comparison between different magnet materials,  we define an equivalent thermal coefficient for the NdFeB magnet to quantify its behavior. The discussion in this paper is based on the assumption of 1\,kg mass realization. Although the NdFeB magnet has a large temperature coefficient of $-1 \times 10^{-3}$/K, it also has a higher coercivity than the SmCo magnet adopted by NIST-4 system. Taking the two versions of the Tsinghua tabletop Kibble balance magnet system \cite{li2024magnet} as an example, the magnetic field in the air gap is 0.45\,T and 0.59\,T for SmCo and NdFeB, respectively. An increase in the magnetic field can reduce the current through the coil when weighing the 1\,kg mass, thereby lowering the coil ohmic power. Thus, an equivalent thermal coefficient for the NdFeB magnet is defined as $\alpha_\mathrm{m}\mathrm{(NdFeB)} = -1\times 10^{-3}/\mathrm{K} \times \left(\frac{0.45}{0.59}\right)^2 \approx -5.8\times10^{-4}/\mathrm{K}$. While the 'equivalent thermal coefficient' is reduced to 60\,\% of the original value for NdFeB, it is still twice as large as the thermal coefficient for SmCo. In the following text, this coefficient is used for calculating the thermal-magnetic error of NdFeB magnets.

\subsection{Two-mode two-phase measurement scheme}
\label{section 3.1}

Figure \ref{fig:tmtp} presents the thermal-magnetic effect analysis with different magnet system sizes, i.e. different $p$ values, $p=1, 0.8, 0.6, 0.4, 0.25$, when the TMTP measurement scheme is used. The calculations are performed using a 2D $rz$-axisymmetric FEA model in air, with the initial system temperature set to 25\,$^\circ$C. As shown in Figure \ref{fig:02} (a), in a full repeatable measurement period $\tau_{\mathrm{c}}=7200\,$s, both weighing and velocity measurement time are set to 1\,hour. During the weighing measurement, the current ramping time (from 0 to $I$ or $-I$) is $\tau_{\rm{r}}=45$\,s, followed by a stabilization period of $\tau_{\mathrm{s}}=90\,$s, with the final 60\,s of data used for analysis. The measurement sequence includes ten mass-on and ten mass-off measurements. The current is kept constant during the stabilization period of weighing phases, and the ohmic power of the coil is set to $p^{-3} \times 5.5\,\rm{mW}$ ~\cite{Li_MagneticUncertainties}, where $5.5\,\rm{mW}$ is the coil ohmic power of NIST-4 system \cite{NISTInvitedArticle2016}. In the velocity measurement phase, the time required for one up or down measurement is set to 90\,s (5\,s for acceleration or deceleration, and 80\,s for data collection), with the central 60\,s used as the final result. The number of up/down measurement pairs is set to 20. 

\begin{figure*}[h]
	\centering
	\includegraphics[width=1.1\columnwidth]{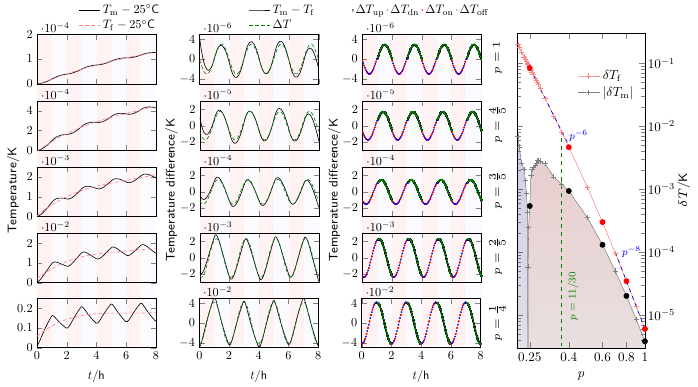}
	\caption{Thermal-magnetic effect analysis of TMTP for different magnet system sizes. The five rows (top to bottom) correspond to the results for $p=1, 0.8, 0.6, 0.4$, and 0.25. The first column shows the change in magnet temperature from an equilibrium state at 25\,$^\circ$C. The second column shows the temperature variation after subtraction of the slow drift together with a fit, $\Delta T$, to a sinusoidal curve. The third column marks the weighing and velocity measurement points for the fit. The right plot shows the maximum periodic magnet temperature variation due to the measurement sequence, $\delta T_{\rm{f}}$ and the absolute value of the averaged magnet temperature change between velocity and weighing measurements $|\delta T_{\rm{m}}|$, as a function of $p$. $\delta T_\mathrm{m}$ is positive in the red-shaded region and negative in the {blue-shaded} region. For $p_0\approx0.2477$, the thermal-magnetic effect is zero. {The five red and five black solid dots represent the five cases ($p$ from $\frac{1}{4}$ to $1$) shown in the left subplots.}}
	\label{fig:tmtp}
\end{figure*}

The first column of Figure \ref{fig:tmtp} shows the temperature change of the permanent magnet, $T_\mathrm{m}$, as a function of time for four measurement periods. By using a low-order polynomial fit, $T_\mathrm{f}$, the overall drift (smooth part which is not coherent with the measurement phase, shown in Figure \ref{fig:01}) can be removed, and the periodic variation, $T_\mathrm{m}-T_\mathrm{f}$, is shown in the second column of Figure \ref{fig:tmtp}. The order of the polynomial fit is the smallest order for which periodic temperature variations can be obtained after subtracting the fit from the data -- usually an order between six and nine. After the subtraction of the polynomial fit, a clear periodic change of $T_\mathrm{m}-T_\mathrm{f}$ can be observed, and since weighing and velocity measurements take the same time, the data can be fit to a sinusoid. The best sinusoidal fit, $\Delta T$, is shown in each subplot, as well as its corresponding measuring points in the weighing and velocity phase, namely $\Delta T_\mathrm{up}$, $\Delta T_\mathrm{dn}$, $\Delta T_\mathrm{on}$, and $\Delta T_\mathrm{off}$. In the last column of Figure \ref{fig:tmtp}, the red curve presents the peak-peak amplitude of the sinusoidal fit, defined as $\delta T_\mathrm{f}$, as a function of $p$. It follows approximately $\delta T_\mathrm{f}\propto p^{-8}$ at $p=1$ and roughly $\delta T_\mathrm{f}\propto p^{-6}$ at $p=0.25$. For the NIST-4 system ($p=1$), the amplitude of the periodic temperature variation $\delta T_\mathrm{f}$ is about 6.0\,$\upmu$K. With the SmCo magnet $\alpha_\mathrm{m}=-3\times10^{-4}$/K, the magnetic field varies $1.8\times10^{-9}$ relatively peak-peak. Take the magnet system of the Tsinghua tabletop Kibble balance as an example: $p=11/30\approx0.37$, the temperature change is 8.1\,mK peak to peak. Using the NdFeB magnet with the equivalent thermal coefficient $\alpha_\mathrm{m}\approx-5.8\times10^{-4}$/K, the relative magnetic field variation is $4.7\times10^{-6}$, which is not acceptable for the $10^{-8}$ level relative measurement uncertainty target.

In reality, the systematic effect introduced by the coil ohmic heating is lower due to the average effect considering data collection. As shown in the third column of Figure \ref{fig:tmtp}, the thermal change between two measurement phases can be estimated as
\begin{equation}
	\delta T_\mathrm{m}=\frac{\bar{\Delta T_\mathrm{up}}+\bar{\Delta T_\mathrm{dn}}}{2}-\frac{\bar{\Delta T_\mathrm{on}}+\bar{\Delta T_\mathrm{off}}}{2},
\end{equation}
where $\bar{\Delta T_\mathrm{up}}$ and $\bar{\Delta T_\mathrm{dn}}$ denote the average $\Delta T$ value respectively for the upward and downward velocity measurement points, $\bar{\Delta T_\mathrm{on}}$, $\bar{\Delta T_\mathrm{off}}$ the average values for mass-on and mass-off weighing measurements.

The black curve in the right plot of Figure \ref{fig:tmtp} shows the change in the absolute value of the measured temperature variation, $|\delta T_\mathrm{m}|$, as a function of $p$, which 
allows a more intuitive analysis of the magnitude of thermal-magnetic errors. Overall, the $|\delta T_\mathrm{m}|$ curve remains well below the $\delta T_\mathrm{f}$ curve, and the reduction from $\delta T_\mathrm{f}$ to $\delta T_\mathrm{m}$ is nonlinear, largely due to the delayed heat transfer. 
When $p=1$, the time delay $\tau_{\mathrm{d}}$ for heat transfer is approximately $1800\,$s, 1/4 of a single measurement period $\tau_{\mathrm{c}}$. In this scenario, the temperature during the velocity measurement is consistently higher than during the weighing measurement, representing the worst case for averaging out temperature variations. At this point, the calculation yields $\delta T_\mathrm{m} = 3.8\,\upmu$K. Using the SmCo magnet introduces a relative systematic bias of $\epsilon=1-(Bl)_\mathrm{w}/(Bl)_\mathrm{v}=\alpha_{\rm{m}}\delta T_\mathrm{m} \approx -1.1 \times 10^{-9}$. 
As $p$ decreases, specifically in the range $0.25 < p < 1$, the temperature variation  $\delta T_\mathrm{f}$ increases due to the significantly reduced heat capacity of the magnet system, but the time delay of the heating $\tau_{\mathrm{d}}$ decreases, enhancing the averaging effect. In this range, the periodic temperature variation is significantly averaged out, and the thermal-magnetic effect may even decrease when the time delay $\tau_{\mathrm{d}}$ becomes sufficiently small. The time delay $\tau_{\mathrm{d}}$ is found to be zero at $p = p_0 \approx 0.2447$, where the thermal-magnetic effect becomes independent of coil ohmic heating and thus reaches its minimum value. Surprisingly, the zero-delay point is not at $p \rightarrow 0$. 

Although this zero thermal-magnetic effect design is promising, the effect is highly sensitive to $p$ at this point $p_0$, limiting its practicality. Without an experimental detection and adjustment method, the technique remains insufficiently robust. 
For $p < p_0$, the absolute value of both temperature variation $\delta T_{\rm{f}}$ and time delay $|\tau_{\mathrm{d}}|$ increase as the magnet size decreases, leading to a sharp rise in the thermal-magnetic error.

For the Tsinghua tabletop Kibble balance, $p=11/30$, the averaging significantly reduces the temperature difference, i.e. $\delta T_{\rm{m}}/\delta T_{\rm{f}}$=0.15, and a temperature
change between two measurement phases, $\delta T_\mathrm{m}=1.2\,$mK, is obtained. With $\alpha_\mathrm{m}\approx-5.8\times10^{-4}$/K, the relative thermal-magnetic error would be $-7.0\times10^{-7}$ using NdFeB magnet. Therefore, using the TMTP scheme makes it difficult to suppress this effect for the Tsinghua magnet system. 

\subsection{Traditional one-mode two-phase measurement scheme}

\begin{figure*}[tp!]
	\centering
	\includegraphics[width=1.1\columnwidth]{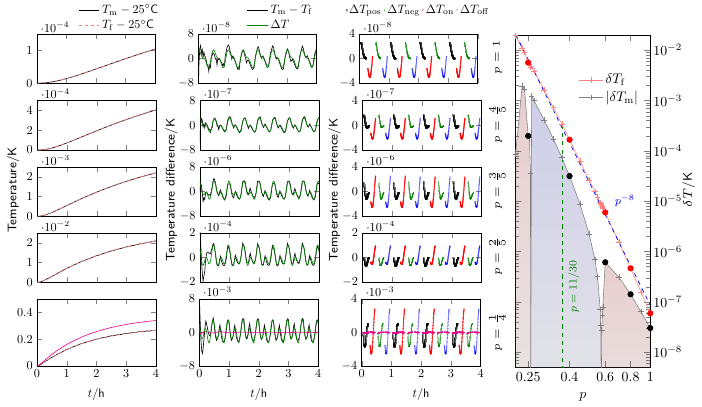}
	\caption{Thermal-magnetic effect analysis of OMTP under different magnet system sizes. The first column shows the magnet temperature change, starting from 25\,$^\circ$C. The second column shows the temperature variation after the slow drift is removed, and the fit, $\Delta T$, is a two-sinusoidal-component curve. The third column lists the weighing and velocity measurement points. The right plot shows the maximum periodic magnet temperature variation due to the measurement sequence, $\delta T_{\rm{f}}$, and the absolute value of the mean magnet temperature change between velocity and weighing measurements $|\delta T_{\rm{m}}|$,as a function of $p$. $\delta T_\mathrm{m}$ is positive in the red-shaded region and negative in the {blue-shaded} region. {The five red and five black solid dots represent the five cases ($p$ from $\frac{1}{4}$ to $1$) shown in the left subplots.} The magenta curves presented in the last row are the result of the proposed measurement scheme with $p=0.25$, whose $|\delta T_{\rm{m}}|$ is reduced by about 2 magnitudes compared to OMTP. }
	\label{fig:omtp}
\end{figure*}

Figure~\ref{fig:omtp} presents the thermal-magnetic effect results under the conventional OMTP measurement scheme. In this example, the current ramping sequence of the bifilar coils is shown in Figure \ref{fig:02} (b). The full measurement period $\tau_{\mathrm{c}}$ is set to 1\,hour, with each weighing and velocity measurement lasting 15\,minutes,
in which $\tau_{\rm{r}}=150\,\rm{s}$ is for current ramping (from 0 to $\pm I$, or from $\pm I$ to 0) and $\tau_{\rm{s}}=600\,$s for a stabilization period during measurement. In each weighing measurement, the first 30\,s is used for mass exchange and control stabilization, and the final 570\,s is for measurement and data collection. For the velocity phase, every stabilization period contains 15 pairs of up and down movements, and the time for one up or down movement is set to 20\,s, where 5\,s for acceleration or deceleration, and the central 10\,s for measurement. 

As discussed above, the OMTP measurement scheme uses a bifilar coil, unlike the single coil used in the TMTP scheme. When adopting the same magnet system with a fixed wiring space, a smaller coil wire gauge is required for winding the bifilar coils to ensure that the $Bl$ factor remains optimal~\cite{Li_MagneticUncertainties}. In this case, the number of coil turns is the same for both schemes, but the cross-sectional area of the wire in the TMTP is double that of the OMTP. This results in a doubled coil resistance for OMTP and, consequently, doubled coil ohmic power (assuming the same current to maintain the force). Therefore, in the FEA calculations, the coil ohmic power is set to $p^{-3} \times 11.0\,\rm{mW}$ for the conventional OMTP measurement.

As observed in the first column of Figure \ref{fig:omtp}, the periodic temperature variation is much smoother compared to the TMTP scheme. Since the symmetry between heating and non-heating in the input power is broken, the temperature variation shown in the second column of Figure \ref{fig:omtp} cannot be adequately modeled by a single-frequency sinusoidal fit. However, we found a sum of of two sines with one half the frequency of the other can effectively describe the differential data. We fit $T_\mathrm{m} - T_\mathrm{f}$ to
\begin{equation}
	\Delta T = a\sin\left(\frac{2\pi}{\tau_1}t + \phi_1\right) + b\sin\left(\frac{\pi}{\tau_1}t + \phi_2\right) + c,
\end{equation}
where $\tau_1$ is the minimum period of coil ohmic power variation, set to $\tau_{\rm{s}}+2\tau_{\rm{r}}=$ 15\,min in this case, and $\phi_1$ and $\phi_2$ are the phase shifts of the two components, while $a$, $b$, and $c$ are constant coefficients. From the second column of Figure \ref{fig:omtp}, it can be seen that the proposed fit $\Delta T$ accurately captures the temperature variation of the magnet, $T_\mathrm{m} - T_\mathrm{f}$. Similarly, the collected measurement data of different measurement phases is shown in the third column, while the fourth plot of Figure \ref{fig:omtp} illustrates the peak-to-peak temperature variation $\delta T_\mathrm{f}$, and the absolute value of the average temperature difference between the two measurement phases $|\delta T_\mathrm{m}|$ as a function of $p$. For the periodic temperature variation's amplitude $\delta T_\mathrm{f}$, the dependence on $p$ follows approximately $p^{-8}$. For the OMTP measurement scheme shown in Figure \ref{fig:02} (b), the temperature difference between the velocity and weighing phases $\delta T_\mathrm{m}$ is obtained by
\begin{equation}
	\delta T_\mathrm{m}=\frac{\bar{\Delta T_\mathrm{pos}}+\bar{\Delta T_\mathrm{neg}}}{2}-\frac{\bar{\Delta T_\mathrm{on}}+\bar{\Delta T_\mathrm{off}}}{2},
\end{equation}
where $\bar{\Delta T_\mathrm{pos}}$ and $\bar{\Delta T_\mathrm{neg}}$ denote the average $\Delta T$ value for the Velocity$(I_+)$ and Velocity$(I_-)$ measurement points, $\bar{\Delta T_\mathrm{on}}$, $\bar{\Delta T_\mathrm{off}}$ the average values for Weighing$(I_+)$ and Weighing$(I_-)$ measurement data. 

Taking the NIST-4 magnet ($p=1$) as an example, the peak-to-peak temperature variation $\delta T_\mathrm{f}$ in the OMTP scheme is $0.06\,\upmu$K, which is one hundred times lower than that in the TMTP scheme. Further averaging the weighing and velocity points can reduce the temperature change associated with systematic effects, resulting in a reduction to $\delta T_\mathrm{m}=0.03\,\upmu$K, corresponding to a relative measurement error of $-9.0 \times 10^{-12}$ when using SmCo magnets. For the Tsinghua tabletop system ($p = 11/30$), when the conventional OMTP scheme is used, the peak-to-peak temperature variation $\delta T_\mathrm{f}$ is 0.34\,mK, and after averaging, $\delta T_\mathrm{m}=T_\mathrm{v} - T_\mathrm{w} \approx -0.07$\,mK. If NdFeB is used, this will result in a relative systematic error of approximately $4.3 \times 10^{-8}$, which remains significant. 

The thermal-magnetic error $\epsilon=\alpha_{\rm{m}} \delta T_{\rm{m}}$ in the OMTP measurement is proportional to the current ramping time $\tau_{\rm{r}}$. Although reducing the current ramping time is beneficial, the duration is unavoidable due to the mass exchange required for mass-on and mass-off measurements.

\subsection{Proposed one-mode two-phase measurement scheme with current ramping}

Lastly, an FEA simulation of the proposed measurement scheme is performed. The setup is the same as in the conventional OMTP scheme, except for the use of the current ramping described by formulas (\ref{i1_v1_eq})-(\ref{i2_v1_eq}). As expected, the result is very smooth, with no noticeable periodic temperature changes related to the measurement sequence. Here, we show the result for $p=0.25$, in addition to the last row of Figure \ref{fig:omtp} as the magenta curves. The temperature variation $\delta T_{\rm{f}}$ and the averaged result $\delta T_{\rm{m}}$, compared to the conventional OMTP scheme, are reduced by about 2 magnitudes, showing a significant impression on the thermal-magnetic error. Note that in the FEA calculations, the exact error reduction of $\delta T_\mathrm{f}$ and $\delta T_\mathrm{m}$, as a function of $p$, is no longer reliable due to calculation noise. 

\section{Experimental verification}
\label{section4}
Here we report an experimental investigation of the thermal-magnet effect on a tiny BIPM-type magnet system. The goal is to observe the thermal changes under different measurement schemes and to validate the FEA calculations. Figure \ref{fig:expsetup} shows the experimental setup: A BIPM-type magnet is used, with an outer diameter of 40\,mm and a height of 42\,mm. The total weight of the magnet is approximately 295\,g. A bifilar-coil with a mean radius of 12.5\,mm is employed, with a total of 500 turns for each coil (wire diameter 0.10\,mm). The resistance of each coil is $R_1=R_2\approx110\,\Omega$. The bifilar-coil is fixed within the air gap of the magnet.

\begin{figure}[tp!]
	\centering
	\includegraphics[width=0.8\columnwidth]{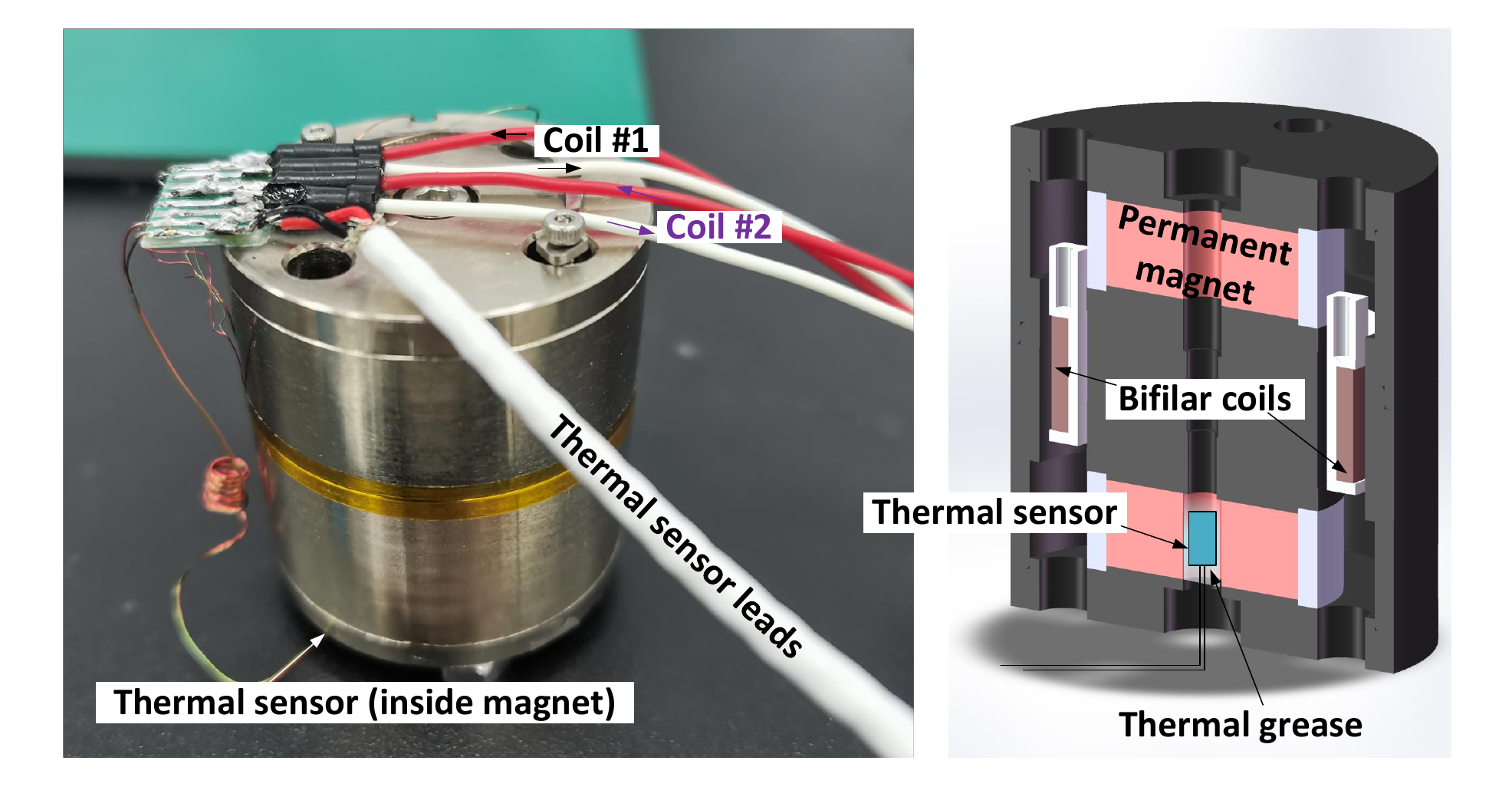}
	\caption{Setup of the experimental test.}
	\label{fig:expsetup}
\end{figure}

The currents in the two coils, $i_1$ and $i_2$, are generated by two commercial Keithley 6221\footnote{Certain equipment, instruments, software, or materials are identified in this paper in order to specify the experimental procedure adequately. Such identification is not intended to imply recommendation or endorsement of any product or service by Tsinghua University and National Institute of Standards and Technology (NIST), nor is it intended to imply that the materials or equipment identified are necessarily the best available for the purpose.\label{fn:NIST_commercial_ disclaimer}} current sources. The current ramping time is set to $\tau_{\rm{r}}=180$\,s (from 0 to $\pm I$, or from $\pm I$ to 0), and the stabilization time for data collection $\tau_{\rm{s}}$ is 600\,s. With a maximum current of $\pm80$\,mA, the maximum ohmic power for each coil is 0.704\,W. A thermal sensor (platinum resistance temperature sensor, PT100 M222 1/3B\footref{fn:NIST_commercial_ disclaimer}) is inserted through the center hole to the position of the permanent magnet, with thermal grease applied to improve contact between the sensor and the magnet. The thermal sensor's output is read by a Keysight 3458A\footref{fn:NIST_commercial_ disclaimer} digital multimeter, 8.5 Digit. 

With the test setup, we investigate the temperature change for three different scenarios: (a) current on and off as it would occur in the TMTP scheme, (b) traditional current switching for the OMTP scheme, and (c) the proposed constant power switching of the current for OMTP mode.  The measurements are summarized in Figure \ref{fig:expresult}. The top row shows the current in the coil(s) and the associated ohmic power. { The temperature reading from the thermal sensor for the three cases is presented in the middle row.} And on the bottom is shown the temperature difference, i.e., which is obtained by removing the mean and the slow varying drift from the middle graphs. 

\begin{figure*}[tp!]
	\centering
	\includegraphics[width=1.1\columnwidth]{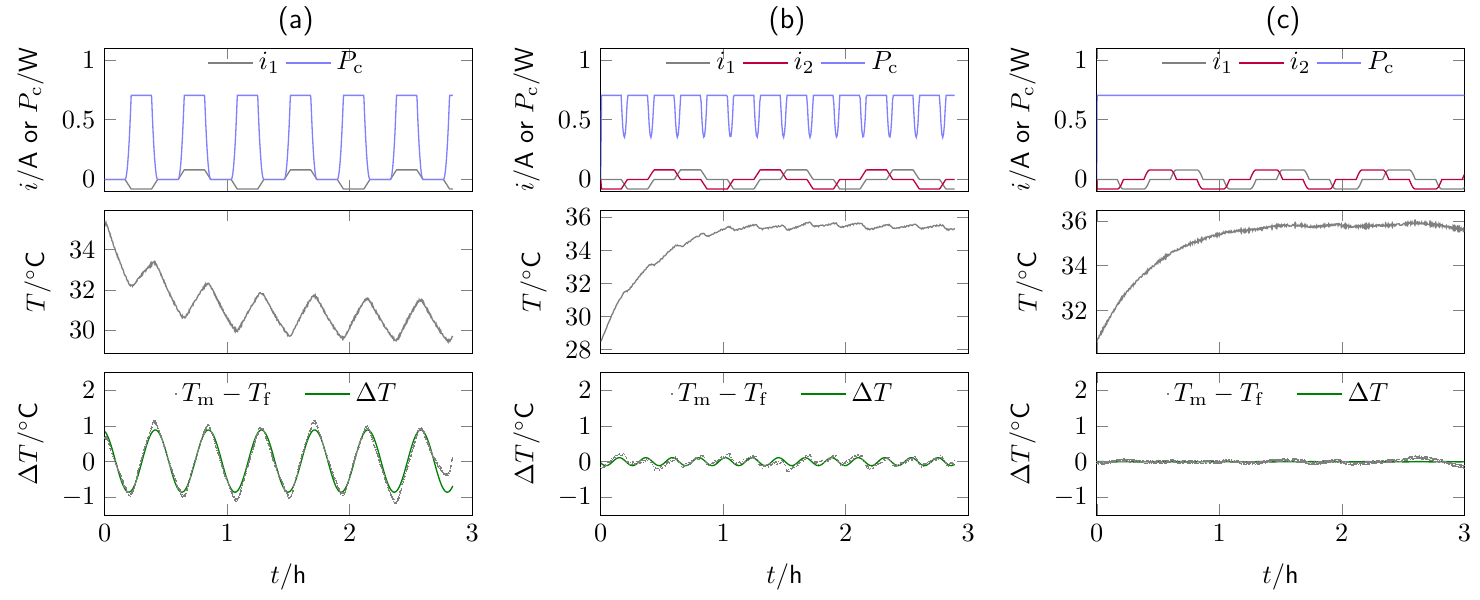}
	\caption{Results of experimental tests under different measurement schemes are shown. Subplots (a), (b), and (c) correspond to the TMTP scheme, conventional OMTP scheme, and the proposed measurement scheme, respectively. The first row displays the currents applied to the bifilar coil and the total ohmic power as a function of time. The second row shows the temperature change of the magnet as measured by the thermal sensor. It is important to note that the initial temperature depends on when the measurement is started since the experiment is conducted in an air-conditioned room. For instance, in (a), the magnet is still in the cooling process, {and its initial temperature is higher.} The third row presents the periodic temperature variations associated with the power inputs with the slow drift removed.}
	\label{fig:expresult}
\end{figure*}

It is important to note that since the experiment is conducted in an air-conditioned room, the initial temperature and its drift slope are random, { depending on whether the magnet temperature has cooled completely to room temperature, but this does not affect the result of $\delta T_{\mathrm{f}}$ or $\delta T_{\mathrm{m}}$.} After removing the slow drift, the periodic variation $T_{\rm{m}}-T_{\rm{f}}$ in the magnet temperature is observed, as predicted. A two-component sinusoidal fit $\Delta T$ shows that the peak-to-peak temperature variations $\delta T_{\rm{f}}$ for the three cases are 1.59\,K, 0.23\,K, and 0.04\,K respectively, and the averaged temperature between two phases $\delta T_{\rm{m}}$ are -0.45\,K, -0.03\,K and 0.004\,K. The FEA calculation of the experimental setup yields temperature variations $\delta T_{\rm{f}}$ of 1.79\,K, 0.20\,K, and 0.001\,K for the three cases, and $\delta T_{\rm{m}}$ are -0.89\,K, -0.01\,K, and $-0.001$\,K, showing reasonable agreement with the experimental result. 

This experimental result is limited by the measurement accuracy of the thermal sensor {($\approx1$\,mK for measuring temperature change)}, but it confirms that the proposed ohmic power compensation method can reduce the thermal-magnetic error by more than two orders of magnitude compared to the traditional TMTP scheme, and by at least an order of magnitude compared to the conventional OMTP scheme in a Kibble balance experiment.

\section{Discussions}
\label{section5}

The principles of the thermal-magnetic effect of a Kibble balance, particularly the TMTP and conventional OMTP schemes, as well as the effectiveness of the proposed approach to eliminate it, have been verified through FEA numerical simulations and experimental tests. Note that the FEA simulation setup, especially the measurement period $\tau_\mathrm{c}$, is referred to the NIST-4 and the BIPM Kibble balance experiments. 
It is interesting to compare the thermal effects of TMTP and OMTP with the same measurement period $\tau_\mathrm{c}$. Here we also calculated the thermal-magnetic effect of TMTP with $\tau_{\mathrm{c}}=3600\,$s when keeping the current ramping strategy unchanged. The result is presented in Figure \ref{fig:tmtp_timedelay} (a), and for the same magnet size $p$, the temperature variation $\delta T_{\mathrm{f}}$, $|\delta T_{\mathrm{m}}|$ of $\tau_{\mathrm{c}}=3600\,$s is a few times lower than that of $\tau_{\mathrm{c}}=7200\,$s, which would not affect much the conclusion obtained in section \ref{section3}.

\begin{figure*}[tp!]
	\centering    
	\includegraphics[width=0.7\columnwidth]{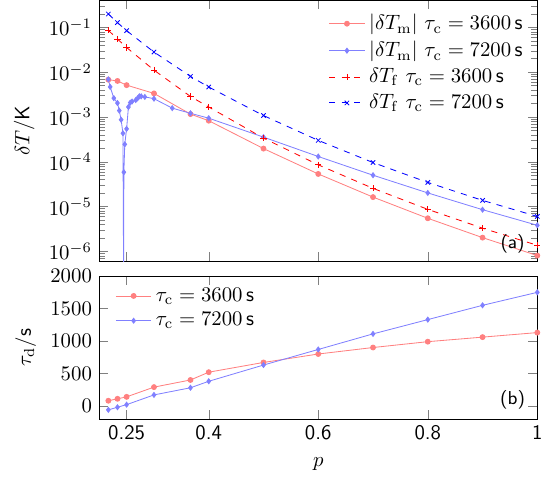}
	\caption{(a) is a comparison of the thermal-magnetic effect for TMTP measurement scheme with measurement period, $\tau_{\mathrm{c}}$ respectively, 3600\,s or 7200\,s. (b) presents the time delay $\tau_{\mathrm{d}}$ as a function of $p$ under $\tau_{\mathrm{c}}=3600$\,s and 7200\,s.}
	\label{fig:tmtp_timedelay}
\end{figure*}

To understand the zero-thermal-effect point, $p_0$, the time delay between the periodic temperature variation $\Delta T$ and the coil heating cycle, $\tau_\mathrm{d}$, as a function of $p$ is also calculated, and the result is shown in Figure~\ref{fig:tmtp_timedelay} (b). The $\tau_\mathrm{d}(p)$ curve with $\tau_\mathrm{c}=7200$\,s clear shows a zero crossing point at $p_0$, where the heating cycle precisely aligns to the temperature variation, allowing a good suppression of the thermal effect. It can also be seen from the comparison of two curves in  Figure~\ref{fig:tmtp_timedelay} (b) that there $p_0$ is not a fixed position and varies with the measurement period $\tau_\mathrm{c}$. A possible explanation of this phenomenon is the heating interaction may be related to the match of $\tau_\mathrm{c}$ and the thermal time constant of the magnet. The detailed mechanism, as seen by some FEA calculations, is complex and beyond the scope of this paper.

The second issue is that in the proposed OMTP measurement scheme, the simultaneous current ramping introduces an overshoot in the Lorentz force. The Lorentz force on coil \#1 and coil \#2 acts either in the same or opposite directions, i.e.,
$F(t) = i_1(t)Bl \pm i_2(t)Bl$, depending on the direction of coil currents. The amplitude of the total Lorentz force can be expressed as
\begin{eqnarray}
	|F(t)|&=&|BlI\left[\pm\sin\left(\frac{\pi}{2\tau_{\rm{r}}}(t - t_0)\right) \pm \cos\left(\frac{\pi}{2\tau_{\rm{r}}}(t - t_0)\right)\right]|\nonumber\\
	&\leq&\sqrt{2}Bl|I|,
	\label{eq:force}
\end{eqnarray}
where the equality holds when $t = {\tau_{\rm{r}}}/{2} + t_0$, resulting in a 40\,\% overshoot in force compared to the conventional OMTP measurement scheme. This overshoot occurs regardless of whether the two coils' Lorentz forces are in the same or opposite directions, so it can be considered a drawback of the proposed scheme. In practice, experimenters can reduce the current ramping speed and allow for a gentler landing on the mechanical stop to protect the flexure hinge.

\begin{figure}[tp!]
	\centering    \includegraphics[width=0.8\columnwidth]{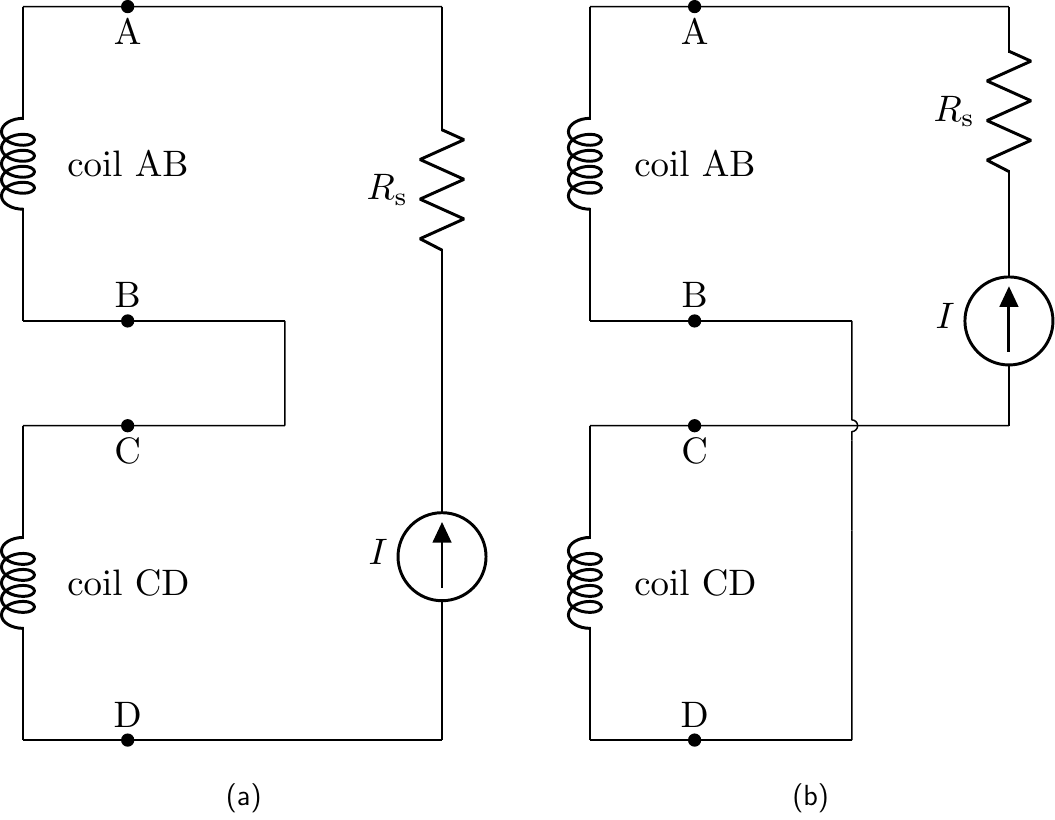}
	\caption{A proposed new circuit topology of the TMTP scheme for the elimination of thermal-magnetic effects and the Lorentz force overshoot. 
		The coil in the air gap is composed of the coil AB and the coil CD, which can be switched for series or anti-series connection. $R_{\mathrm{s}}$ is the sampling resistance and $I$ represents the current source. (a) is for the weighing phase, and the series connection is applied. (b) shows that the two halves of the coil are anti-series connected in the velocity phase.}
	\label{fig:TMTPcircuit}
\end{figure}

Considering the elimination of both the thermal-magnetic effect and the overshoot in the Lorentz force, a new circuit topology of the TMTP measurement scheme is proposed as Figure \ref{fig:TMTPcircuit}, which makes it possible to allow the coil current passing through both the weighing and the velocity phase. The coil in the air gap consists of two coils wound together with all four terminals, AB and CD, made available to the outside, which can for example be achieved with bifilar winding. The coils can be connected in series (Figure \ref{fig:TMTPcircuit} (a)) or in anti-series (Figure \ref{fig:TMTPcircuit} (b)) using a switch box. In the weighing phase, two coils are in the series connection, which acts the same as the traditional TMTP measurement scheme. While in the velocity phase, the coil AB and the coil CD are anti-series connected as displayed in Figure \ref{fig:TMTPcircuit} (b), and the current through the coil maintains the same magnitude as the weighing phase, so 
it is possible to realize the elimination of thermal-magnetic effects while making two Lorentz forces cancel each other out in opposite directions. The voltage drops across the two coils during the velocity phase are
\begin{equation}
	U_{\mathrm{AB}}=IR_{\mathrm{AB}}+(Bl)_{\mathrm{AB}}v\;\;\mbox{and}
	\label{Uab_AB}
\end{equation}
\begin{equation}
	U_{\mathrm{CD}}=-IR_{\mathrm{CD}}+(Bl)_{\mathrm{CD}}v,
	\label{Ucd_CD}
\end{equation}
where $R_{\mathrm{AB}}$,$R_{\mathrm{CD}}$, $(Bl)_{\mathrm{AB}}$ and $(Bl)_{\mathrm{CD}}$ represent the resistance and the magnetic geometrical factor of the coil AB and the coil CD in the velocity phase. When the parameter consistency of the two coils is well maintained, and the parameters drift against time, the temperature is also the same, $R_{\mathrm{AB}}=R_{\mathrm{CD}}$
,$(Bl)_{\mathrm{AB}}=(Bl)_{\mathrm{CD}}$ can be guaranteed. The magnetic geometric factor $(Bl)_{\mathrm{v}}$ can be obtained as
\begin{equation}
	(Bl)_{\mathrm{v}}=(Bl)_{\mathrm{AB}}+(Bl)_{\mathrm{CD}}=\frac{U_{\mathrm{AB}}+U_{\mathrm{CD}}}{v}.
	\label{newBL_v}
\end{equation}

The feasibility of the proposed new TMTP circuit topology is promising, though it requires higher consistency of coil parameters, and a more complex switch box needs to be designed for coil series switching and voltage measurement. While the wiring is similar to the OMTP scheme, the advantage is that all the copper is energized during the weighing phase, i.e., for the same amount of copper and current, the proposed four-terminal TMTP coil produces twice the $Bl$ value, thus twice the Lorentz force for mass calibration.

\begin{figure}[tp!]
	\centering    \includegraphics[width=0.8\columnwidth]{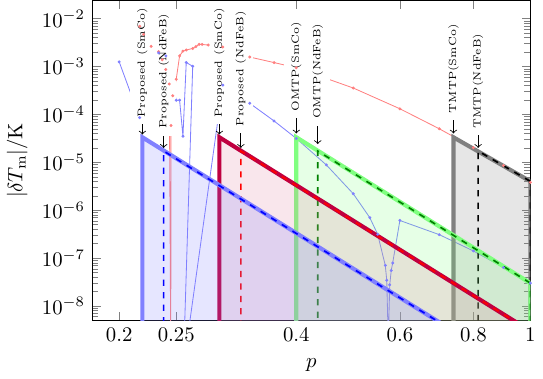}
	\caption{The magnet size boundaries for achieving a target Kibble balance relative measurement accuracy of $1\times10^{-8}$ are illustrated. From right to left, the gray, green, red, and blue shaded areas represent the thermal-magnetic error boundaries for the traditional TMTP, the conventional OMTP, and the proposed schemes (with error suppression of one and two magnitudes), respectively. The solid and dashed lines correspond to the use of SmCo and NdFeB magnets. The data points in red and blue are $|\delta T_{\mathrm{m}}|$ of the TMTP and the conventional OMTP measurement schemes respectively, as a function of $p$.}
	\label{fig:07}
\end{figure}

The analysis of the thermal-magnetic effect across different measurement schemes enables the determination of size boundaries to achieve a target Kibble balance measurement accuracy. Figure \ref{fig:07} provides an example based on the mean thermal change, $|\delta T_\mathrm{m}|$, obtained in section \ref{section3}. By limiting the relative thermal-magnetic error to $1\times10^{-8}$, the magnet size range for different measurement schemes can be identified. In the shaded region of Figure \ref{fig:07}, the solid and dashed boundaries correspond to the use of SmCo and NdFeB magnets, respectively. 

It is important to note that, in principle, the proposed ohmic power compensation method can effectively eliminate the thermal-magnetic error regardless of the magnet size. However, to account for imperfect cancellations and ensure reliability, an improvement of at least one order of magnitude and, at most, two orders of magnitude compared to the conventional OMTP measurement scheme is considered for the proposed approach. These one-magnitude and two-magnitude boundaries are indicated in red and blue, respectively, in Figure \ref{fig:07}. Table \ref{tab:01} lists the minimum size and weight of the magnet system for each measurement scheme.

\begin{table}[tp!]
	\caption{The minimum size and weight of the magnet with different measurement schemes}
	\label{tab:01}
	\centering
	\begin{tabular}{l|ccc}
		\hline
		Measurement scheme & $p$  &  $\Phi_\mathrm{m}$/mm & $M$/kg\\
		\hline
		Reference magnet (NIST-4) & 1 & 600 & 850\\
		\hline
		TMTP (SmCo) &0.740 &444.0 &344.4 \\
		TMTP (NdFeB) &0.815 &489.0 &460.1 \\
		\hline
		OMTP (SmCo) &0.400 &240.0 &54.4 \\
		OMTP (NdFeB) &0.435 &261.0 &70.0 \\
		\hline
		Proposed $\frac{1}{10}$OMTP (SmCo) &0.296 &177.6 &22.0\\
		Proposed $\frac{1}{10}$OMTP (NdFeB) &0.322 &193.2 &28.4\\
		\hline
		Proposed $\frac{1}{100}$OMTP (SmCo) &0.219 &131.4 &8.9\\
		Proposed $\frac{1}{100}$OMTP (NdFeB) &0.238 &142.8 &11.5\\
		\hline
	\end{tabular}
\end{table}

It is worth noting that optimizing the magnetic circuit design can further extend the boundaries shown in Figure \ref{fig:07}. For instance, in the Tsinghua tabletop magnet system, the fringe field at both ends of the air gap is compensated by reshaping the inner yoke boundary~\cite{li2024magnet}, resulting in a significantly larger uniform magnet field range compared to a conventional BIPM-type magnet design. This improvement in the uniform field range allows for more coil windings, leading to a larger $Bl$ value. To maintain a constant magnetic force, the coil ohmic power can be reduced. In the Tsinghua system, $Bl\approx400$\,Tm, the coil resistance for each coil is about 400\,$\Omega$, and the currents needed for calibrating a $1\,\rm{kg}$ mass are $\pm 12.5\,\rm{mA}$. The actual coil heating power is 62.5\,mW, which is a substantial reduction compared to the power scaled from the NIST-4 system (when $p=11/30$, $P_\mathrm{c}$ is $129.8$\,mW for the NdFeB magent system using the bifilar coils).

\section{Conclusion}
\label{section6}

In conclusion, this study addresses the significant challenge of thermal-magnetic errors in tabletop Kibble balances by proposing a novel measurement scheme that precisely regulates coil ohmic power, and analyses the mechanism of thermal-magnetic effects in different measurement schemes. 
Through comprehensive FEA simulations and experimental validation, the proposed method demonstrates its effectiveness in significantly reducing temperature variations during measurements. Specifically, the new scheme reduces thermal-magnetic errors by over two orders of magnitude compared to the conventional TMTP scheme and about one order of magnitude compared to the traditional OMTP scheme. The results confirm that maintaining a constant coil ohmic power minimizes temperature fluctuations, thus enhancing measurement accuracy and breaking down the limitations on the minimum size of tabletop Kibble balances.

{The FEA simulations and experimental validation presented in this paper were conducted in air. However, since high-precision Kibble balances may operate in a low-vacuum environment ($<0.01$\,Pa), the thermal-magnetic effects could differ slightly due to the influence of thermal radiation. This aspect warrants further investigation in future studies.} 
While the proposed approach shows great promise, it also presents challenges, such as the need for precise current control and a good switch scheme design. %
Further research is required to refine the technique for practical applications in high-precision, compact Kibble balances.

\section*{Acknowledgment}

Shisong Li and Weibo Liu would like to acknowledge the financial support from the National Natural Science Foundation of China (No. 52377011).

\end{flushleft}

\end{document}